\documentclass[aps,prd,showpacs,nofootinbib,preprint,tightenlines]{revtex4}
\def\be{\nopagebreak[3]\begin{equation}}
\def\ee{\end{equation}}
\def\ba{\nopagebreak[3]\begin{eqnarray}}
\def\ea{\end{eqnarray}}

\newcommand{\teta}{\rlap{\lower2ex\hbox{$\,\tilde{}$}}\eta{}}

\begin{document}
\title{Unspeakables and the Epistemological path towards  Quantum Gravity.}

\author{ 
 Daniel Sudarsky\\ \small
 {\it Departamento de Gravitaci\'on y 
Teor\'\i a de Campos, Instituto de Ciencias Nucleares} \\ \small
{\it Universidad Nacional Aut\'onoma de M\'exico} \\ \small {\it 
Apdo. Postal 70-543 M\'exico 04510 D.F, M\'exico} }

\begin{abstract} 
 We  offer a critical assessment of some generic features of  various of the current approaches
towards the construction of a Theory of Quantum Gravity. We will argue that there is a need for further conceptual clarifications before such an enterprise can  be launched on  a  truly well grounded
 setting, and that one of the guiding principles  that can be viewed as part of the reasons 
 for successes of the past theoretical developments is the identification of Unspeakables: Concepts that should not only play no role in the formulation of the theories, but  
 ones that the  formalism of  the theory itself should 
  prevent from ever been spoken about. 

 \end{abstract}

\maketitle
\section{Motivation}
The search for  a theory that bridges the apparent abyss that separates the General  Theory of Relativity and Quantum Theory has absorbed the efforts of  an important segment of the theoretical physics community  for decades, and although 
there is no doubt, that progress in various  of the lines of approach has been substantial,  particularly 
during the last decade, we are still far from the  stage in which we could argue we have a fully satisfactory theory.  The main contenders  (as measured by the size of the community that  works in the approach) are, of course,  String Theory \cite{ST}  and Loop Quantum Gravity \cite{LQG}, with some less popular candidates  like  The Partial Ordered Causal Sets program \cite{POSET}, and  the  Non-commutative Geometry program \cite{NQG}  maintaining an important presence in the field, while the initial impetus of the so called Wheeler de Witt \cite{WDW}approach is almost completely gone.

I find it to be a healthy  trend that people have started looking at the limitations and problems of the different approaches  because science can only progress in an environment of robust  and open  criticism, and  of constant focus on the problematic aspects of its current understandings. 
In this regard,  several works have been written concerning the limitations of String Theory \cite{STProb}, among which we can cite  its reliance of the  formulation of  the theory on a  background metric for space-time,  the absence of a clear and {\it a priori} identification of the  physical degrees of freedom, the absence of a unique vacuum, etc. The Loop Quantum Gravity approach suffers from the absence 
of a wide enough class of  solutions to the Hamiltonian constraint, the existence of  ambiguities that might be similar to those associated with the non-renormalizability of the perturbative approach to the quantization of gravitation \cite{Ale}, to name the most important. 
This is not to say, however, that some of the above programs have  not encountered some important successes.
In fact, one of the main challenges that the theory
of quantum gravity is supposed to address is the calculation, from first principles,  of the  
Beckenstein-Hawking entropy for black holes.  In this regard, both String Theory and Loop Quantum Gravity  can claim relative success in  a limited  set  of circumstances \cite{QBH}.  Technical difficulties impair their progress into the more  general setting where they can be  expected to be able to deliver similar success. Enormous amounts of efforts are being invested in these enterprises.
 However, I should like to point out that there is a realm where they are not expected to be able to deliver satisfactory answers and that concerns precisely the type of situation  where strong quantum effects are combined with strong gravitational effects: The so called  Schr\"odinger  Black Hole, a situation where a quantum process determines whether or not a Black hole is present at an earlier time \cite{SchBH}.  The situation is essentially a setup where an initially static thin spherically symmetric shell with mass M  will be  conditionally allowed to collapse at the speed of light  leading to the formation of a Schwarzschild Black Hole. The crucial aspect is that the condition for allowing the collapse depends on the outcome of a quantum mechanical measurement, set up in such a way that  there will be regions of space-time  that will be located behind the event  horizon  in the event that the collapse is triggered, but that will  
 precede in  time the event at which the quantum measurement takes place. The  result is that the location, and,  therefore, the area of  the event horizon will be undetermined at some times, leading to a problem when trying to assign to it  the corresponding  entropy.  The example is quite  contrived, of course,  but it serves to illustrate several fundamental aspects of the current understanding about black hole entropy and its identification with the area of the event horizon, that one would expect a fundamental theory would be able to elucidate. 

 The relative small amount of interest by the quantum gravity community on this, and similar issues, is an illustration of what I feel is an overemphasis on the technical  aspects of quantum gravity in detriment of a concerted effort in advancement on the  conceptual problems that plague the unification of the theories
 \cite{Conceptual}.

%[OUT] With the possible exception of the POSET program goal of QG: status and the lack of conceptual %progress.

 My point of view is that true success will elude us unless we start focussing on the various conceptual problems that  should face any attempt to construct a unified version of quantum theory and gravitation, if that enterprise is to lead to a satisfactory theoretical understanding of physical reality.
     By these, I mean, among other things, facing the measurement problem in quantum theory \cite{QMProblem}, in particular, in regard with the suggestions that some novel aspects of physics would be  needed to address the issues in a satisfactory manner \cite{NewPhys}. These issues can be seen to become particularly acute when dealing with the problems that arise when trying to extend the standard interpretative schemes to the realm of quantum field theory\cite{Sorkin}, and when  considering questions concerning quantum theory and cosmology \cite{Us}.  There are, of course,  several other  problems that need addressing such as  the problem of time  in quantum gravity \cite{time-QG}. In this regard I  would like to echo the concerns expressed,  for instance, by R. Penrose \cite{Penrose} but  I might add, that in my view,  what is at stake is more than what it seems, because  if we did manage to  solve the problem of quantum gravity  without having to address the foundational problems of Quantum Mechanics,  we would be in a situation in which all hope for a guiding path towards its solution would be gone. Therefore, the  posture  I am exposing here could  be considered to arise out of  a certain ``faith"  that nature would not deprive us from reaching a true understanding of the rules that it follows.  Faith that can  only be justified by the previous historical development of our science, and perhaps by an irrational preconception about the innate abilities  of men and the “un-mischievous” character of nature itself.

 In this article, I will argue in favor of a certain type of guiding principle, which has the advantage of allowing itself to  being casted in negative terms,  thus avoiding the need for a clear  and {\it a priori} elucidation of the ideas. The principle is based on the recognition of unspeakables.  That is, the identification of concepts that are  being dragged uncritically from previous stages of the understanding of the phenomena, and that  upon further analysis  are  understood  to be of such nature that they should play  no role whatsoever in the more profound theoretical construct one is seeking.
 The principle to be followed once the unspeakables are identified, is that the theory should be constructed to avoid even the possibility that such concepts might be considered within its  formalism. 

\section{The role of unspeakables in the past conceptual developments  in theoretical physics}

  Looking back at the history of the development of our current physical theories it is easy  to spot several occasions where the breakthroughs were associated with the identification of previously unexpected classes of unspeakables.  Let us briefly review some of the most important cases.

1)  The Galilean insight about  the   absence of meaning in the concept of absolute rest frames in mechanics: One of the first instances, if not the first one, in the history of science where a great leap forward was achieved in connection with the rejection due to lack of meaning  of a previously well accepted concept involves the   Aristotelian notion of absolute state of motion. It is well known that in the Aristotelian ideas  about nature, the everyday intuition of rest and motion was elevated to an absolute conceptual construct thought to be central in the laws governing the behavior of objects. This was  particularly   explicit as it concerned inert objects, but in  truth ultimately thought to apply  to  all objects, whose motion would need to be caused by some external entity.
  
  This notion when confronted with the observation  that the celestial bodies  were,  in contrast with the terrestrial ones, in a state of permanent motion, led to the  even more pernicious notion  (for the development of science) of a separation between the ``Laws of the Earth" and the ``Laws of the Heavens".  Needless is to recount here the details of the story that goes from Ptolomeo, Copernicus, etc. to Galileo who faces for the first time, in a coherent and methodical way, the concept of absolute motion. In doing so, he is able to unearth the unexpected compatibility of the idea that the Earth might be moving through space (around the sun) and that we might not feel the motion. The story of Galileo’s description of the happenings on the surface of a ship as described by the sailors aboard it and by the onlookers on the ground, brought to life the centrality of the  notion of relative motion in contrast with the concept of absolute motion of Aristotelian physics, which, as we know, led to the Newtonian revolution, mother of other scientific and technological revolutions itself.
 One way, and the one I am advocating here, to look at the rise of  the centrality of the  notion of relative motion is to consider it the natural outcome of taking absolute motion and making it an Unspeakable. That is, in all subsequent developments, thinkers about these issues would have  known  that they were deviating from the correct path whenever, in their arguments, they were making use, even in indirect ways, of the notion of absolute motion.  The emergence of such concept would be automatically taken as a sign that  something was not being analyzed correctly\footnote{Nowadays,  notions  of absolute  motion  are again being considered
 \cite{LIV},  but in those cases people  regard space-time as endowed with some additional  microscopic structure, and the absolute  motion  under consideration is, in fact,  motion relative to  a preferential rest frame associated with such  hypothetical structure. Such point of  view, thus evades the  above criticisms of the older conceptual incarnation of absolute motion, but interestingly enough  such ideas have encountered  exceedingly tight bounds for the effects\cite{LIVBounds}  and  serious problems when analyzed  in the light of modern quantum field theory \cite{LIVProb}.}.
 In fact, at the time  when Newtonian Physics was being  applauded for all its successes, there were some voices pointing out some unsettling aspects of the conceptual construct:  The existence of absolute space, with respect to which no absolute  notion of uniform motion could be identified, but which  permitted the identification of absolute acceleration.  The theory  of Newton, and, in particular, the famous three laws of motion, were  supposed to be valid  descriptions of physics if that description was made using observations carried out in connection with any {\it inertial } frame.
 Those frames, which  played such a central role both in Newtonian Mechanics, and latter in Special Relativity, were  identified as  a problematic concept by E. Mach who observed that the notion of absolute acceleration without reference to any particular object seem to be quite different from the other conceptual constructs of the theory. The idea that an object  would be  accelerating with respect to space itself was deemed epistemologically ungrounded as there seem to be no possible way to  directly observe  this space, and thus, of determining if the object was or not accelerated with respect to it\footnote{In fact, it is quite amusing to note that even today this issue can put in serious trouble  many  undergraduate and  graduate physics students and even some otherwise  very well versed   and qualified professionals in our  field:  Q: What is the law of inertia?   A:  A free particle  moves with constant velocity. Q: Is that supposed to hold in any  possible frame?  A: No!,  of course not,  only in  inertial frames!  Q:  What is an ``Inertial Frame"? A:  Well...one in which the law of inertia holds. Q:  So the first law is not a law at all, but a mere  tautology, right? A:  Yes.. well no.. hmm.. let me think.  There was something about the frame of the fixed stars, and those moving with uniform velocity with respect to it......but there are no fixed stars!  Hmm...It must be  the frame associated with CMB, right?.
   The answer is of course the following:  We take 3 free point particles (the identification of which is another story, but which in particular  implies that we must  neglect gravitation as  its universality would preempt the existence of free particles) whose motion is not coplanar, adjust our motion so that we see them moving inertially (i.e with constant velocity), then we { \it define} our frame to be an inertial one.  The law of inertia then  states,  that {\it all other}  free particles  would  move inertially when described in that frame. This is clearly not a tautology as one can imagine a universe where this would not  hold.  Anyway the point here is just an illustration of the type of considerations that I feel have been largely  forgotten in our days. I think that  the fact that this  type of problem is  so common, is a result of  an educating system that puts to much of its emphasis  in ``training people to solve problems" in detriment  of  the  work associated with  clarifying conceptual structures.}.

 2)   As we all know that  Galilean revolution came to be revisited as a result of the studies on the electromagnetic phenomena. Those studies culminated  in the formulation of   Maxwell´s equations,  which,  in turn,  predicted a finite and specific  value for the velocity of light. The conflict  of such notion with the Galilean relativity of velocities,  led naturally to the concept of a special reference frame, the only one,  where Maxwell's equations where thought to be  exactly valid:  The frame of the ``Luminous Ether". These notions led in turn,  to the search for indications of the Earth´s motion relative to such frame, and eventually  to the development of the Special Theory of Relativity.
  While it  is possible to look at this development as emphasizing the absolute nature of the velocity of light,  I think it is more illuminating to regard it  not only as  reenforcing  the Galilean identification of
  the unspeakable nature of the concept of absolute motion,  but  as the unearthing of a whole new collection of unspeakables: The duration of a process, the notion of simultaneity, the  length of an object, etc.  Of course, these concepts would resurge in a more modest incarnation, as relative concepts to be associated with a particular frame of reference, but with no meaning whatsoever in the absence of that specification. The theory, on the other hand, leads to some new concepts that are thought  to have  absolute meaning:  The four momentum of a particle or of a system, the interval between two events etc.
 We should, however, note  that the special theory of relativity, made absolutely  no progress in dealing with one of the major criticisms of the Newtonian conceptual edifice:  The epistemologically problematic  concept of  absolute state of acceleration and the related notion of inertial frames. 
  
3) The next revolutionary change in our conceptual understanding in physics emerges from the
 impulse  to bring  gravitation into the new general  scheme of special relativity. As is well known,  this culminated in the development of the General Theory of Relativity. Here, the  geniality of Einstein is observed, among other things, in his willingness to give up further conceptual pillars of the theoretical understandings of the day, just to free the theory from 
 notions that had little epistemological justification, and, in particular,  to address at the same time
 the issue of making the theory compatible with new insights gained from  special relativity and
 the  epistemological problems associated with  the concept of inertia, as pointed out by Mach.
 
I consider the following to be a convenient way to look at the development of General Relativity, whether or not it is an accurate historical account of that process (most probably not). Let us take the notion of inertia as described in the footnote 1.  As noted there, the conceptual construct  would work if we were able to turn gravity off.  But of course  we are not, so why is  such construction of any utility whatsoever?
 Why is it that we can make so much progress using this notion to do physics, to construct buildings and bridges, fly airplanes etc? After all, the Earth is certainly not inertial, and furthermore there is gravity all around us.  It is true, we always consider gravity in our treatments,  but we, in practice, consider only that due to Earth and ignore,  for instance, the gravitational effects associated  with the  Sun or the Galaxy.
  The answer lies, of course, on  the universality of gravity, which can be seen as indicating that one may, in carrying out the ``construction" of the inertial frame described in the footnote  2,  ignore 
  the fact  that gravity is there,   and proceed  with the prescription as if gravity  did not exist.  The universality of free fall  would ensure that when following the instructions given in  footnote 2,  all  the bodies,  including the observer would ``fall at the same rate", and thus that the  law of inertial will be satisfied in the constructed frame.
  The only point is that construction  will only have local validity:   the law of inertia will be valid to the extent that you limit observations to a sufficiently  small  space-time region.
  The notion of global Inertial frame is thus turned into an Unspeakable, and in so doing, help us understand the effectiveness of the Laws of Newtonian physics as used in our everyday lives.
  In short,  while it seemed that gravity would invalidate the whole conceptual construct leading to  the clarification of the content of the Law of Inertia, one crucial characteristic of gravity,  namely,  its universality,  saves the day, and resurrects the notion of inertia but in a  fashion which ties it inextricably with the Equivalence Principle.  In this regard, the view, which I think is quite common outside the gravity community, that the Equivalence Principle  is  just a ``curious" aspect of gravitation,  is quite misleading.  
  As  we saw,  we could  not even  make sense of the laws of inertia  if the gravitational interaction  
   did not lead to a universality of free fall,  but still,  acted on all bodies, and  thus  denied  us
    the free particles we need to define inertial frames. Thus,   the Equivalence Principle should be seen as   the ``reason" behind inertia itself. 
  
  This stage  of  understanding evidently gives rise to other unspeakables such as the notion of  ``a global inertial frame", and all related ones, but perhaps, even  more unexpectedly it renders generally meaningless,  concepts such as  the total energy or four momentum of an extended system: There is,  in general, no way, to give meaning to  the sum of  the four momenta of a collection of particles, or the integral of the energy momentum  of a
 continuous distribution of matter fields.   To see this, consider a region of space-time in which we have a distribution of matter characterized by an energy momentum tensor $T_{ab}$, and imagine that we want to compute the total energy momentum tensor.  First, of course, we would need to identify an instant of time, which in general relativity would correspond to identifying a certain 3 dimensional space-like hypersurface $\Sigma$. The problem is then how to define $\int_\Sigma T_{ab} (x) dV_x$, as we do not know how to add tensors that are associated with different points. The possibility of using the notion of parallel transport is precluded, in general, due to the  dependence of the resulting transport with the path that is used and that is characteristic of curved space-times. 

 In fact this type of problems leads to other unspeakables such as the generic notion of center of mass of a distribution of matter, and to the  absence of a well defined procedure for the identification of a world line characterizing an extended system in general \cite{Dixon}.  We will see that this particular unspeakable reappears  in a more dramatic way at the moment we start to consider  the concepts associated  with gravitation in a quantum context.

4)  The quantum revolution, exemplified  for simplicity in the  relatively simple   theory of
Quantum Mechanics, brings about some well known unspeakables.
We learn that  there is no meaning to the  notion  a particular  path to a particle, as a path would require the particle to have at every instant a well defined position and  a well defined momentum. In fact, we learn that there is no physical reality to the points in phase space, and no meaning for that concept exists in the formalism. Two important new unspeakables. In fact, in general, things can be even more  problematic because a  subsystem that is part of a larger system  can not be thought as
having a well defined state of its own: The situation where there is entanglement imply that 
 the parts  can  have, at most, a partial description  (in terms of a density matrix) where other relevant information will appear only in the complete  description of the system as a whole \cite{Mermin}.
  There is, of course, an extended literature dealing with the surprising and counter-intuitive aspects of the quantum world, where, in fact, the notion of unspeakable seems to first find its use\cite{QW}, so here I only mentioned some of the most important concepts  of the classical theories that became, strictly speaking, unspeakables after the quantum aspects of nature are taken into account. 
 
5)  Quantum field theory in curved space-time. The marriage of quantum theory with special relativity brings about the quantum theory of fields. There one learns  of complications that have to do with incorporating the negative energy solutions, the ensuing notion of anti-particles, and, in dealing with all but the simplest case on noninteracting theories the  complex issues of renormalization, anomalies etc.  However, it is only at the next stage when one is forced to consider the  situation resulting from replacing the Minkowski space-time background with the more general  case of  a curved space-time background, that an  important  actual new unspeakable emerges: The concept of particle. For the benefit of readers not familiar with this point  we present next  a very brief description of the problem.

 Consider  a theory of a  scalar quantum field in a background space-time with metric $g_{\mu\nu} $ and described by the Lagrangian density
  ${\cal  L} =\sqrt{-g} ( \nabla _\mu \phi \nabla _\nu \phi  g^{\mu\nu} -m^2  \phi ^2)  $.  In constructing the quantum theory we are led to write the field as an operator 
  $\hat \phi  (x) = \Sigma_i  ( \hat a_i f_i(x) +\hat a_i^\dagger f_i(x)^* )$, where  the functions $f_i (x),  f_i (x)^*$ are a complete set of, suitable normalized,  solutions of the Klein Gordon equation in the background space-time, i.e. 
  $\nabla^\mu  \nabla _\mu f_i (x)  -m^2  f_i(x) =0 $, and the $ \hat a_i , \hat a_i^\dagger $ are annihilation and creation operators,  respectively.   It is at this stage that the vacuum state  $ |0>$ can be defined by the requirement that $\hat a_i |0'> =0,   \forall  i $.  The point is, that we could have used  another  complete set of solutions to the Klein Gordon equation, for instance, if we define 
  $g_i(x) =  \alpha f_i (x) +\beta  f_i (x)^*$,  with suitable conditions on the coefficients  $\alpha$ and $  \beta$ to preserve the  appropriate normalization,  the set of functions $g_i (x),  g_i (x)^*$ are an equally  valid   complete set  of solutions leading now to the decomposition of the field operator as
  $\hat \phi  (x) = \Sigma_i  ( \hat b_i g_i(x) +\hat b_i^\dagger g_i(x)^* )$.  The new  operator coefficients
   $ \hat b_i , \hat b_i^\dagger $ are  again annihilation and creation operators in the sense that they satisfy the corresponding commutation relations. In fact, one can check that 
   $\hat b_i  = \gamma \hat a_i  +\delta \hat b_i^\dagger $ for some coefficients $\gamma$ and $\delta $  ( depending on  $\alpha$ and $  \beta$). Then we can define a new vacuum state $|0'> $ 
   such that $\hat b_i |0'> =0,   \forall  i $. It is quite clear that the two 
   vacuua are  very different states, because 
   $\hat b_i |0> =  (\gamma \hat a_i  +\delta \hat a_i^\dagger )|0> = \delta \hat a_i^\dagger |0> \not=0$.
   This problem is not encountered in the usual quantum field theory construction  in Minkowski space  because in that case  one can use  the fact that  Minowski space-time is stationary, to define the notion of positive/negative energy solutions. The ambiguity is thus resolved,  by the  election to associate the annihilation operators with the positive  energy solutions.
  However,  in dealing with non stationary space-times, there is, in general, no  way to make any sort of canonical selection of the mode functions that are used in the expansion of the quantum field operator.  For a more comprehensive discussion of these issues see for instance \cite{Wald}. Thus  the notion of  a vacuum  state is not well  defined, i.e.  it becomes a new unspeakable, and the notion of a particle, being the result of acting by a creation operator on the vacuum, becomes,  in the absence of  further specifications, another unspeakable.

\section{New  unspeakables in Quantum Gravity }

 One thing that needs emphasizing is that almost without exceptions the concepts that are identified as unspeakables at  one stage of understanding of nature, maintain their characters of  unspeakables    in the  later developments associated with  a deeper understanding.  As indicated above, one such exception is the velocity of light that becomes absolute as a result of  developments  that occur after 
 the stage in which all velocities are deemed to be relative.
  
 Therefore, in discussing the next (and perhaps, ultimate) level of understanding of nature, which  we naturally associate with  a theory of quantum gravity, a  set  of guiding lampposts should be that in its conceptual development all previously identified unspeakables should be regarded as such. 
  Namely, that theory should have no place, and make no use of concepts like: ¨ {\it  the energy of a particle} ( unless one specifies  the frame with respect to which it is considered) on account of 1) \& 2),
   {\it  the notion of a particle}  other that as a relatively  compact distribution of energy momentum, on account of 5),  and  then {\it the notion of  the energy- momentum of such distribution}, on account of 3).
  
I find it quite surprising, therefore, that  some of the attempts to formulate a quantum theory of gravity,
 have no qualms in embarking  in such an enterprise, carrying along  some of the  conceptual baggage that should have been  discarded according to the lessons from the previous levels of understanding.
 
 I will leave it to the people working on such approaches to  address this point, and will instead   focus  now on further lessons that, I believe, can  be of use in the search for the desired theoretical path, one that should have, by its non- reliance on unspeakables,  the right  {\it a priori } features giving one the confidence that it might end in a success.  
 
 Thus, I propose we ask ourselves next:
What about the marriage of Quantum Theory and 
gravitation?. Are there any new unspeakables that can be unearthed even before we have a theory of quantum gravity? For, if some of those were to exist, their identification could be an important guiding light into the eventual formulation of the theory. If we  did identify  one unspeakable, and it turned out to be a concept of common and widespread use in the  discussions about Quantum Gravity,  the efforts  that would be required to avoid its usage,  would be, in themselves efforts to replace the unsuitable concepts by others having more potential for lying at the basis of an appropriate formulation of the sought theory. 
In the following, I will discuss, using an epistemologically inspired  line of reasoning,
what I believe should be regarded as some of the new unspeakables that we must learn to contend with. Needless is to say that as the theory must contain, or at least accommodate, in principle, the previously  discussed and well established   theoretical advances,  all the unspeakables listed before should also be unspeakables of quantum gravity: {\it  energy,  length,   absolute velocity, precise number of particles in a state, etc}\footnote{One of the astonishing facts that can be observed in looking at some of the prevailing approaches, such as String Theory,  is the complete naturality with which some of the well known  unspeakables, throughly identified in the preceding stages of the theoretical advancement, come back into the lexicon of the discussion. The most glaring example of this is  the concept of energy.}. Our  discussion will concentrate in what we believe to be  the novel unspeakables. In this discussion, we will allow ourselves  the use of some of the concepts and language of the pervious stages, because the discussion,  carried out  in the absence of  a  selected candidate  theory of quantum gravity, will be carried out  in the “meta-language” afforded by the  earlier stages. It is on the actual  candidate theories that the demand should  be imposed that their formalism is such that it  makes it impossible to talk about the unspeakables.

  Let us consider the notion of space-time geometry. In  all preceding stages of development of this notion, and, in particular, in The General  Theory of Relativity,  the space-time geometry is directly linked to observations: Using  free test point particles one can  identify   the space-time geodesics and from these, in principle, one can extract the geometry \footnote{In practice, there are of course other issues like those tied  to coordinate choices and the closely related gauge conditions, that would need to be resolved  in actually presenting the object $g_{ab}$), but we want to focus on aspects connected  with the physical objects themselves rather than their particular mathematical  representations.}.  In fact, it should be clear  that one can, in principle,  read off  the metric  of space-time, for instance,  by identifying both the light cones and the conformal factor, and  that in a classical realm there are no impediments to doing so.  The point  in which we want to focus is that, in order to give  an operational meaning to the geometry, we need  those free test point  particles. The problems one  faces in going to the quantum gravity realm will be, precisely, those associated to the  notions represented by the  four words: free, test, point, particles.
Some considerations of similar nature have been expressed before for instance in \cite{Alhuwalia:dd}.

i) The first issue one confronts is the identification of objects as free, and  it  involves their coupling to other objects. The point is, of course, that nature does not offer us in reality any interaction free matter field.
Retaking for a moment the  notion of particles, we know that all of them do interact via some of the interactions  in standard model. Photons scatter of  other photons and even neutrinos are not free from the electro-weak couplings.  This is, of course, irrelevant at the macroscopic scales where one is usually  concerned with  gravity, but  if we pretend to extrapolate the conceptual constructs to the sub-atomic realm,  which presumably are  still quite removed from the 
regime where quantum gravity would be relevant, we must face the fact that there, the situation is fully reversed. In those conditions all actual particles and fields must be accepted as fully interacting.
In fact,  if  there did exist some species of particle that was actually  free, it would be rather difficult if not outright impossible to use them to identify the geodesics as they would be essentially  invisible, and, in the situation where they could be considered  as acting only as test objects,  they would be actually invisible.  We could, however, take the view that this is only an issue ``in  practice" and not one of principle, but even then  this would not be the end of our problems.

ii)  Let us take, for instance, the view that in the limit, when electrically neutral particles,  are infinitely  far from each other, they can be regarded as free. Consider then, the task of identifying,  for instance the null geodesics of a space-time,  which would lead us naturally to  consider  photons as the ideal
 class of  objects to be employed for the task. The problem we would need to confront is 
 that,  even if distant from all objects,  the interaction of the photon with the virtual particles  
of other species, results in the failure of the photon to propagate along null geodesics\cite{Photons}.
   In fact,  things become even more dire when  we note that a  real photon is not actually in a four  momentum eigenstate, because if it were, that would imply that it is not at all spatially  localized,  a condition rendering  it  completely useless for a world line identification in any case. The point is that  in these circumstances  its four momentum is not null.
  Let  us consider the flat  space-time limit and  the decomposition of the photon state into four momentum eigenstates reflected in a distribution of four momenta represented by 
   $P(p^\mu) = f(p^\mu)\delta(p^2 )$.  Then, the mean  four momenta is
     $ <p^{\mu}> = \int d^3p P(\vec p) p^\mu $   which is, in general,  a time-like  vector,
     as can be seen by noting that
     $  <p^{\mu}>  <p_{\mu}>  = \int d^3p d^3p'   P(\vec p)P(\vec p' ) p^\mu p'_\mu $ and
      $ p^\mu p'_\mu = E E' -  \vec p  . { \vec p} \ ' = E E' ( 1 -  Cos(\theta) ) \geq 0$,   where $\theta$ is the angle between $  \vec p $ and  $\vec p \ ' $ with the  last equality holding only for $\theta =0$. Thus, if we  want to consider  partially localized  photon wave packets, we would not be investigating the null geodesics.

   iii)  Moreover, having recognized that we must deal with  photons that are extended objects  with a four momentum that is
   not well defined,  we are then led to   consider  the  ``mean momenta" and ``mean position" to
     be the quantities which in the quantum world  would correspond to the classical quantities that are used in the identification of geodesics.  The problem is that, in a curved space-time, it is rather unclear how to make sense of  such concepts. We can see the severity of the  difficulty  by considering
      the simpler case of  a classically extended object, described, say, by its  energy momentum tensor $T_{ab}(x)$:  We already noted   the problem of extracting, on a given hypersurface  the four momentum of the  object,  but we would have to face  what seems to be an even more daunting problem: that  of extracting on that hypersurface the center of mass, and, in general, the world line of the center of mass. 
      The classical (in contrast with the quantum mechanical) version of the problem has been considered 
      in various works in the past \cite{Dixon}, but the  results are not sufficiently reassuring, even in this relatively simple  setting (which suggests that the problems would become even more serious in the full quantum setting).  To name some of the disquieting aspects, let  us start by noting that the constructions considered so far  for defining the center of mass world line of an extended object in curved space-time are all based on the corresponding definition in Spacial Relativity, and, there, the only possibility which satisfies the essential requirements of frame independence and four momentum conservation, leads  surprisingly to the non-vanishing of the Poisson bracket among the center of mass coordinates \cite{Price}. This would undoubtedly translate in non-commutativity in the quantum version of the problem. The next problem is that, in the  curved space-time setting, the definitions can be made only as long as the extent of the matter distribution is small enough so that the energy momentum tensor of  the object in question vanishes outside a convex normal neighborhood  (a neighborhood $U$ where any two points $p,q \in U$ are connected by a unique geodesic within $U$).
        Other problems are that, in general, the four momentum of the system is not tangent to the  center of mass world-line, and perhaps, more disastrous,  that its world line is not,  in general,   a geodesic of the space-time geometry.
        In short, in the quantum realm we have, in principle, no way to  access, or identify, the  geodesics of space-time. This means that, even if we consider a situation where the metric is well defined, we would  have  no way to access it.  It is an object that is in essence  unobservable.   So we should ask ourselves: Why should such an object play any role in physics?.
        
        iv)  The possibility of exploring space-time with test particles,  and, in general, test probes, has been considered at length and we will not repeat here all the arguments, but essentially recall the fact that a test object would have to be both highly localized (to explore the small regions of space-time one  would in principle be interested on),  and to have a  sufficiently small contribution to the energy momentum tensor, so that it could be considered to be  testing of the preexisting  metric with a negligible effect on it. The point  is that these two aspects are mutually exclusive when considering probes from the quantum realm. These considerations seem quite valid of course, but they are a bit different  from some of the other considerations  we have made above, in that the former  rely  on the dynamics of General Relativity in reaching the conflicting stage.  That is, the considerations made in regard to this last point,  are based on the extrapolation  to untested realms of Einstein's equation  $ G_{ab} = 8\pi G_N T_{ab} $, something that, in principle, one might want to reconsider.  Moreover,  we see that one can take the formal limit $G_N \to 0$  and use it to bypass the problem in at least certain limits.  The  problems  raised in i), ii) and iii)
 indicate that the notion itself, of a space-time metric  even in the limiting regime where $G$ is set to 0, would be inappropriate in the quantum realm.
       
       v) One further obstacle that  would  seem to emerge from the previous considerations is that, as we saw, in curved space-time,  quantum field theory allows, in general no satisfactory definition of a particle.  It seems to me that this is not really an issue as long as one could  consider  a concentrated  bunch of energy momentum tensor.  That is, as long as there are states for which the energy momentum tensor remains localized in sufficiently small regions.  The  problem that remains, of course, is that of extracting from those states the underlying geometry, which as we saw, could  probably not be done relying on the identification of center of mass world lines with the  geodesics  of the space-time.

\section{What is, or should be,   a Quantum Geometry?}

Our experience with other theories describing  fundamental aspects of the physical world, such as Maxwell's Theory or the Standard Model of Particle Physics, is that there is  a relatively well  defined procedure to transform a classical theory into a quantum one.  We call this the quantization of the theory.
This procedure  takes the fundamental objects from the classical theory $\Xi_{Class} $ and
  transforms them into  ``similar" objects   $\hat \Xi_{Quant}$.  The similarity just alluded refers to both 
 their tensor structure ( i.e., the quantum version of  a vector field, is an operator valued vector field), and the type of information they convey. The scheme has been quite successful, of course, but it
 is nonetheless a mysterious procedure  and it is fair to say we do not completely understand 
 what does  it mean,   and why  does it work so well \cite{Quantization}.
 
When  we acknowledge  that,  in this respect, gravity seems to be quite different from other theories,  for instance  in  not yielding after all the heroic efforts made so far, to  direct (or even  rather indirect) application of the quantization recipe,  it becomes, not only legitimate, but also strongly compelling to start considering the issue in a different light.
 
What is being  advocated here, is  a   reconsideration  of what should be the object representing  geometry in a quantum world.  By this I mean what type of information it should be thought to convey,
 and what should it not. Some of these considerations have been expressed in a work that advocates that geometry in a quantum world  should be  an essentially relational construct \cite{CRS}.  That is certainly not the first time that relational  ideas have been advocated, either at the pure quantum mechanical level \cite{Rovelli1},  or in connection with gravitation\cite{relational}, but perhaps  it is the first case in which the proposal is made in conjunction with the suggestion of  doing away with a quantum version of the space-time geometry.  In  \cite{CRS}  a suggestion  is made  regarding a modification of the algebraic approach to quantum field theory in curved space-time to  make it compatible with a novel concept of   quantum space-time \cite{rainer}. Here, we will limit ourselves to discussing the use of the  notion of unspeakables and avoid entering into any detailed proposal realizing the guiding principle.
 
  In considering these problematic issues,
we must keep in mind that there are,
in  principle, two roles played by the physical geometry in a geometric
 theory of gravitation: A) It is  an entity codifying relationships
 between ``events" defined in terms of the ``matter" content of the
 theory (the quotations indicate that these notions would need to be clarified and specified in the appropriate context before any formal developments could  even be attempted), and B) it is a dynamical  entity whose behavior is part of
 the subject of the theory.
 I should focus in aspect  A), which  needless is to say will not be analyzed exhaustively or to any sort of finishing stage.  Aspect B) will necessarily  become susceptible of  a careful analysis only   after A) has been thoroughly carried out.

  Let us commence  by reviewing  the role of geometry in classical (as
 opposed to quantum) physics. More precisely, one  wants to consider  the
 meaning of the   assignment of  a ``geometry" to our physical world. We have
 learned from Einstein that we should be concerned with the  ``measurements"  of
  ``intervals" between ``events". It is quite clear that before  addressing this issue one  must identify these events. In the classical realm,  we could do this,  for instance, by singling
 out one point  on the world line of a point-like object,  something that could  be  
  done, in principle,  by considering the intersection of two  such world
 lines (we  can  think of the ``passing of one  object  in front of a
 particular observer", or the emission of a light pulse from a certain
source). When considering the measurement of the interval,  one should  focus on our reliance on
a  physical device to actually  specify the measuring procedure.  Normally, we think of  using a ``clock", if the interval is time-like;
 a ``ruler",  if the interval is space-like; or define it as null if the two
 events can be connected by a light signal.
In all
 these cases, we must recognize the central role played by  the  physical objects, clocks, rulers, or light pulses,
 in order to  allow us  to determine  what the geometry is.  Here the point is not to restate 
 the obvious fact that we need such objects to measure the geometry.
 What we want to stress is that we need to specify these physical objects
 in order  to {\it define} what we mean by  ``the geometry of our physical
 world".  That is,  we need to  deal with those issues in order  to give meaning to the words.
 It must be noted,  for instance,  that it is, in principle,  not {\it  a priori}
  at all certain, that the geometry could be defined  in such a way that it  does  not depend on the objects we choose to use  in  defining it \cite{Reichenbach}.
  For instance, we could  have chosen  to measure spatial
 intervals (of say, an instant, or  spatial section of space-time  taken for simplicity to be static)
 with either   ceramic rods,  or alternatively  with  metallic rods, and imagine a
 situation in which the temperature  through the region  being examined
 is not uniform.  Under such circumstances, the straightforward
 determination of the geometry of the region  with the two types of roads would give different results.
 For instance, the geometry determined with  the ceramic rods could be
 flat while the one determined with the metallic rods could  have nonzero
 curvature. This is all obvious, and every physicist ``knows"   that the
 use of the ceramic rods is the ``correct choice" in this case because of
 the large thermal  dilation coefficient  of  the metal. Actually, we
 would be told that even if we use ceramic rods we need to correct for
 the residual  thermal effects, and only after doing so,  would we  obtain  a faithful  determination of
 the ``true geometry" corresponding  to that  which would be
 determined by means of ideal ``length preserving"  rods. 
 We all know that there are no
 ideal rods, however, the more critical fact is that,  in principle, there is  a serious  problem in  establishing what is  even meant by a ``length preserving"
 object.  Let us consider how would we know, even in principle, when  we have identified an  object  that does not change its length?.  The evident answer is: by measuring its length in  different
 circumstances.    However, in order to measure its length  in the various circumstances  and compare the results, we need
 to use a length preserving object! We have  found ourselves immersed in  a nasty  circular 
 set of definitions.  Technically   we could consider ---and in practice
 we do--- replacing standard physical objects by ``generalized objects":
 that is  actual physical objects complemented by well defined prescriptions
 of how to correct for the changes in their  assigned lengths under
 specific circumstances (such as taking into account the corresponding
 thermal expansion coefficients and  accompanying every length
 determination with a simultaneous temperature determination).  However, even after doing so we
 are still faced with the issue of  having to define the operational
 procedure to  determine the length preserving  assignment to a physical
  object or a generalized object. The consideration
of other methods, as  for example, the  measurement of  an
 object's length with light signals and clocks, while taking the speed
 of light as 1 by definition,  only converts our  problem into that
 of choosing clocks that run at a ``constant" rate.  It is evident that
 this faces us with complete analogous conundrums:  How could we measure
 the rate of ticking of a clock to determine whether it is ``constant"
 without the use of another clock?
 In classical physics one solves these
 dilemmas by the use of  judicious definitions which  make use of the
 following empirical fact, F: {\it There exist objects and clocks
 (usually generalized objects and clocks) that, when used  to define
 the lengths of intervals, result in empirical laws of physics that
 are particularly simple (as for example, the law of inertia)}
 \cite{Reichenbach}.  In this way, we select the objects,  in practice generalized objects, that give
 an empirical content to our assignment of geometry to the classical
 level of description of our world. 
 
 Let us turn now to the quantum realm, and consider what  are the objects, actually
 the class of objects $\cal O$, that give a meaning to the geometry at
 the quantum level? Does such class of objects exist at all? Are there
 different classes of objects, say  ${\cal O}_1$  and ${\cal O}_2$,  each
 leading to equally simple laws of nature, which  are however different
 for the  different  selections of the class of objects? In this case, we
 would  be confronted  with  the problem of  having two different  geometries
 for the same physical situation, one associated with the class ${\cal O}_1$
 and  the other  with the class ${\cal O}_2$! No attempt to give
  definite answers to  these questions will  even be considered at this point
   as we are only  attempting to raise awareness of  the problematic issues.

 There are several issues associated with the previous considerations that need further discussion.
 First, we note that, while  at the classical level we must
 consider the objects which are used to define the geometry (i.e
 the objects satisfying F) as classical objects, at the quantum
 level,  the natural objects that  should be considered as defining
 the geometry should be themselves ``quantum objects". Thus, given
 that the meaning of  the physical geometry arises  solely from
 statements which must be expressible
  in terms of the objects selected to define it, such meaning should be
 taken to be a certain codification of correlations between physical
 ``events" (this word is used here in an imprecise sense because at
 this point the discussion refers to both the classical and the
 quantum cases) associated with such objects.  At the classical
 level,  we could be  talking about the number of ticks of a clock
 along the world line segment joining two events with which the
 clock  eventually coincides. At the quantum level, we need to
 consider very different sorts of things. One must then expect such
 correlations and their codification to  depend on the type of
 objects one is considering, and,  therefore, the kind of objects
 representing this information (the kind of objects that represent
 geometry) should be expected to be very different in the classical
 and the quantum cases. In particular, we don't expect the quantum
 objects to be described by the same constructs as ordinary quantum
 matter. These differences should be as significant, say, as the difference  between a
 space-time vector (considered here as describing the state of a
classical particle) , and  a wave function (considered here as
 describing a quantum state of a particle).
 Note that up to this point we have not specified what we mean by
``quantum objects''. Our considerations have so far been very general.  
On the other hand, taking a look at the previous advances, it seems clear that the quantum objects will quite likely  be connected with either quantum states of quantum fields, or some further generalization of such conceptual constructs.

   Following the general approach taken in this manuscript, that it is easier to make negative statements than to make concrete proposals,  and to further illuminate the ideas  towards which we are driven
   by the epistemological considerations so far, let us  end  by discussing, what  should {\it not} be done and why. The
 (most widely used) classical description of gravitation involves a space-time (i.e.,  a
 manifold with a  pseudo-Riemannian metric) $(M, g_{ab})$, which
 carries information about the behavior of geodesics, their points
 of intersection,  which would define events, the length of the
 interval between two such events along a  given geodesic, etc. It
 is natural that such an object (i.e. a tensor field) should be used
 to describe the geometry which is defined in terms of point particles
 whose states can be described in terms of a four vector $u^a$ and a
 point $x$ in space-time.
  On the other hand, when the geometry is defined in terms of quantum
 fields\footnote{At this point the use of these words should not be taken
 to indicate the standard mathematical  description of such objects,
 but the underlying physical entities  they are thought to
 represent.}, the physical counterpart of a  geodesic (viewed as
 the world line of a classical test particle) or  the physical
 counterpart of a four vector tangent to a space-time point (viewed
 as a perfectly  localized particle with a well defined four momentum),
  cease to exist! They have turned into unspeakables.  That is, the objects that would give meaning to
 $g_{ab}$ are not part of the realm of physics we are attempting to
 describe. Therefore, we should not try to construct a quantum
 counterpart of $g_{ab}$, i.e., a quantized gravitational field
 $\hat g_{ab}$ or any equivalent object, because such a construct
 would be a quantum object carrying information about classical
 correlations! That is,  the quantum object  would carry information about 
 unspeakables, and such a thing could not be information at all.
 In fact, doing this would have meant that we changed aspect B) of the geometry from classical
 to quantum, but did not change aspect A). Our previous arguments indicated that would make no sense.

Recognizing and keeping in mind these points seem to me essential for
 considerations about the nature of a Quantum Theory of Gravitation.
 The fact is that  the standard views on these
issues rely  on idealizations,  that are not unlike the idealizations
 of pre-relativistic physics about the existence of absolute space and
 absolute time, or those in pre-quantum mechanics about the existence of
 well defined  particle trajectories or the existence of a fully
 deterministic physical world. 
 The fact is that the  starting point of most attempts to  construct candidate  theories  of quantum gravity
  fail to  properly guard  against the use of rather well identified unspeakables.
  As we saw in retrospect,  the failure to properly identified unspeakables and to take 
  precautions against their use, turned out to be  serious impediments
 for  the advancements  in theoretical development in the past.
 I believe that
 the notion of a physical geometry
 existing independently of the physical objects with which to determine
 it (in the sense of defining it), is an impediment towards the
 construction of a quantum theory of gravitation.
  
\section{Conclusions}
 
  We have  argued for an  approach to the question of bringing together Gravitation and Quantum Theory
   that  starts by identifying the concepts that should be ascribed the role of unspeakables of that theory.

  Having made such identifications,  the challenge would be to identify the  concepts that should replace the ones deemed obsolete, and to construct the language and mathematical structure that would, at the same time, allow one to consider the relevant issues, and that  would make it impossible to  deal with, or even to mention the unspeakables.  As an illustration  of what I have in mind, we can point to Quantum Mechanics:  once one decided that in this theory one describes the state of a particle by a wave function one can not, even if one tries, go back and consider a particle with a well defined position and momentum.  The formalism does not allow consideration of such  a thing.  Something similar should be our goal, and  our guiding principle indicates that the formalism for the sought theory, should  be such that the identified unspeakables, can not even be referred too. 
  
Needless is to say that  I have not made any real concrete proposal here that matches the desiderata that has been argued for  throughout  the paper, but it is my hope that the points of view expressed would be  of  help, at  least in identifying the  general character of  the  approaches to the problem,  which based on many of the previous experiences in the historical development of our current understanding of the foundations of physics,  would  seem  to offer an {\it a priori } enhanced likelihood of success.

\section*{Acknowledgments} It is a pleasure to  acknowledge a continuous history of  lengthy interactions with Dr. C. Chryssomalakos, and previous work with Dr. A. Corichi and  Dr. M. Ryan, as well as various discussions with Dr. D. Ahluwalia, that have influenced me in developing the views exposed here.

\noindent  This work was supported
 in part  by DGAPA-UNAM
IN108103 and CONACyT 43914-F grants.

\end{document}